\documentclass[12pt]{article}
\usepackage{amsfonts}
\usepackage{amsmath}
\usepackage{amssymb}
\usepackage{adjustbox}
\usepackage{graphicx}
\graphicspath{{./simuplots/}}
\usepackage[normalem]{ulem}
\usepackage{natbib}
\usepackage{bibentry}
\usepackage{color}
\usepackage{authblk}
\usepackage{array}

\begin{document}

\title{Comments on ``Challenges of cellwise outliers'' by Jakob Raymaekers and Peter J. Rousseeuw}

\author[1]{Claudio Agostinelli}
\affil[1]{Department of Mathematics, University of Trento, Trento, Italy \texttt{claudio.agostinelli@unitn.it}}

\date{14 Febbraio 2024}

\maketitle

\begin{abstract}
The main aim of robust statistics is the development of methods able to cope with the presence of outliers. A new type of outliers, namely ``cellwise'', has garnered considerable attention. The state of the art for dealing with cellwise contamination in different models is presented in \bibentry{raymaekers2023}. Outliers in time series can be treated as cellwise outliers, a further discussion on this subject is presented.

\noindent \textbf{keywords}: Additive Outliers, Cellwise contamination, Innovative Outliers, Propagation of Outliers, Time series.
\end{abstract}

\bibentry{raymaekers2023} is an excellent review on cellwise contamination and related problems in robust statistics. Of particular interest is the discussion regarding the analysis of time series in the presence of outliers (Section 2.7). The construction of robust procedures is a very interesting and challenging problem in this context and methods for cellwise contamination can play an important role in time series analysis. Most estimating procedures for time series are very unstable in presence of outliers and, even though there are several proposals in the literature, a complete satisfactory robust procedure is missing.

There are at least two different types of outliers: additive (AO) and innovative (IO). To illustrate their nature and related problems, a time series $\mathbf{z}_T = {z_1, \ldots, z_T}$ ($T \in \mathbb{N}$) from causal and invertible autoregressive moving average models are considered
, i.e., $\Phi(B) z_t = \Theta(B) a_t$ with $a_t$ being a white noise, and $\Phi(B)$ and $\Theta(B)$ being lag polynomials. A corrupted series $\mathbf{y}_T$ has elements given by
\begin{align*}
y_t & = \sum_{j=1}^n h_j v_j(B) \delta_t(\tau_j) +  z_t  \\
& = \sum_{j \in \text{AO}} h_j \delta_t(\tau_j) + \frac{\Theta(B)}{\Phi(B)} \left(a_t + \sum_{j \in \text{IO}} h_j \delta_t(\tau_j)\right),   \qquad t=1, \ldots, T
\end{align*}
where $h_j$ represents the magnitude of the $j-$th outlier, $\delta_t(\tau)$ equals $1$ if $t=\tau$, and $0$ otherwise, $n$ is the total number of outliers in the time interval $1, \ldots, T$ and the function $v_j(B)=1$ for an AO and $v_j(B) = \left[\Theta(B)/\Phi(B)\right]$ for an IO. From the moving average representation of $y_t$, it is seen that AO affects only the corresponding observation, while each IO propagates according to $\Theta(B)/\Phi(B)$.

Now, let us assume that $y_t$ shares the same representation of $z_t$, that is $\Phi(B) y_t = \Theta(B) r_t$ with respect to the process $r_t$. Then, the relation among $r_t$, $a_t$ and $z_t$ is given by
\begin{align*} \label{equ:rt}
r_t & = \frac{\Phi(B)}{\Theta(B)} y_t = a_t + \sum_{j=1}^n h_j v_j^\ast(B) \delta_t(\tau_j)  \\
& = \frac{\Phi(B)}{\Theta(B)} \left(z_t + \sum_{j \in \text{AO}} h_j \delta_t(\tau_j)\right) + \sum_{j \in \text{IO}} h_j \delta_t(\tau_j), \qquad t=1, \ldots, T
\end{align*}
where $v_j^\ast(B)=v_j(B)^{-1}$. Therefore, based on the autoregressive representation of the process, an IO affects only one error term $r_t$, while each AO affects all subsequent errors according to $v_j^\ast(B)$.

Given a contaminated time series, it is not possible to uniquely identify the outliers generating process, since it is always possible to find several outliers generating processes (maybe with infinite elements) that generate the same contaminated time series. To illustrate this, consider an AR$(1)$ process $z_t = \phi_1 z_{t-1} + a_t$. When an AO occurs at time $\tau$ with magnitude $h_{\tau}^{\text{AO}}$, the observed values are $y_\tau = z_\tau + h_{\tau}^{\text{AO}}$ and $y_t = z_t$ for $t \neq \tau$. However, the same time series can be obtained introducing a pair of IOs at times $\tau$ and $\tau+1$, such that $h_{\tau}^{\text{IO}} = h_{\tau}^{\text{AO}}$ and $h_{\tau+1}^{\text{IO}} = - h_{\tau}^{\text{AO}} \phi_1$.

This contamination process can be described by a white noise process $h_t$ and two stationary Bernoulli processes $A_t$ and $I_t$ with probability of success $p_A$ and $p_I$, respectively, all the processes being uncorrelated. Then, $r_t = a_t + \left(A_t \frac{\Phi(B)}{\Theta(B)} + I_t\right) h_t$. Under the above assumptions, $r_t$ is a white noise.

A robust procedure should be able to provide a parsimonious representation of the outliers and to bound their effects. There are two sources of outliers propagation to take into account. The first one is due to the combination between the type of outliers and the model representation, e.g., additive outliers do not propagate in $y_t$, while they do in $r_t$. The second one concerns the estimation procedure at hand, depending on whether it is based on lagged values of $y_t$, or $r_t$, or both. As an example, let us consider the \citet{hannan1982} estimation procedure, which can be summarized as follows for an ARMA$(p,q)$:
\begin{enumerate}
\item a high order AR$(m)$ with $m > \max(p,q)$ is fitted to the data using the Yule-Walker algorithm; let $\hat{\phi}_{1}^{(m)}, \cdots, \hat{\phi}_{m}^{(m)}$ denote the fitted coefficients;
\item the unknown $r_{t-1}, \cdots, r_{t-q}$ ($t=m+1, \cdots, T$) are estimated by $\hat{r}_t = y_t - \hat{\phi}_{1}^{(m)} y_{t-1} - \cdots - \hat{\phi}_{m}^{(m)} y_{t-m}$;
\item use least squares to estimate the parameters in the linear model
\begin{equation*}
y_t = \phi_{1} y_{t-1} + \ldots + \phi_{p} y_{t-p} + \theta_1 \hat{r}_{t-1} + \ldots + \theta_q \hat{r}_{t-q} + \varepsilon_t, \qquad t=m+1, \cdots, T \ .
\end{equation*}
where the error terms $\varepsilon_t$ obey the classic assumptions.
\end{enumerate}
In this procedure, it is easy to see both sources of propagation in action. Step 1. involves only $\mathbf{y}_T$, hence it is observed propagation of IOs and the presence of cellwise propagation since a (truncated) autoregressive representation is used in the Yule-Walker algorithm. Step 3. suffers from all possible sources of propagation.
The number of contaminated {\it rows}, i.e. containing at least one contaminated cell, can easily be larger than $50\%$. Furthermore, the first source of propagation can be enough to break most of the modern robust methods for cellwise contamination, since this source alone can easily lead to datasets where for all subset of pairs of variables (columns) the number of contaminated rows exceeds $50\%$.

In the following section, a more detailed examination is discussed concerning the reasons and possible remedies of the not satisfactory performance of some classic cellwise regression procedures when used in the time series framework. In the example presented in Section 2.7 of \citet{raymaekers2023}, two cellwise robust methods are used: 2SGS for regression \citep{leung2016} and cellMCD \citep{raymaekers2022}. The 2SGS for regression is a two steps procedures; in a first step a set of filters are used to identify cellwise outliers and flag them as missing values, while in the second step a robust procedure for multivariate location and scatter ables to cope with casewise outliers and missing values is used. cellMCD is instead a method that deals with cellwise and casewise outliers by minimazing one goal function.

It is assumed only the presence of AOs and use the same data generating process as in Section 2.7 of \citet{raymaekers2023}; however every $7-$th entry in the original data $y_t$ is replaced by the outlying value $4$ (rather than $10$), providing an even weaker effect, which is, nevertheless, harder to detect by the procedures. From Table \ref{tab:performance} (first two rows), notice that the cellMCD is slightly degraded, while the 2SGS is more badly affected: this is due to the default filter \citep[UBF-DDC,][]{agostinelli2015,leung2016}, which is not able to flag any contaminated observation. However, there are two simple modifications to improve the performance of the 2SGS. First, apply a univariate filter \citep{gervini2002,saraceno2021} directly on the time series once and then apply the Generalized S-Estimator \citep[GSE,][]{danilov2012} to the flagged data.
This is implemented in function \texttt{ARTSGS} of the supplementary R code. Table \ref{tab:flagged} gives the performance of different flavor of filters in flagging the additive outliers. The univariate Gervini-Yohai filter \citep{gervini2002} with $\alpha=0.95$ (UGY95) detects $32\%$ of AOs, the same performance is obtained with $\alpha=0.6$ (UGY60). This result improves the final performance of the 2SGS estimator (see third row in Table \ref{tab:performance}). The univariate Halfspace filter \citep{saraceno2021} with $\alpha=0.88$ (UHS88) identifies all the AOs with only $7\%$ of false positives, and the final performance is very satisfactory. For comparison, similar results can be obtained by applying a classic S-estimator to the complete cases only (UHS88CC), i.e. to the rows that do not contain any flagged cells by the UHS88 filter (see the last row in Table \ref{tab:performance} and function \texttt{ARS} in the supplementary R code).

\begin{table}
\begin{center}
\begin{tabular}{lrrrr}
\hline
Method  & $\phi_1$ & $\phi_2$ & $\phi_3$ & $\sigma$ \\
\hline
2SGS    & 0.314 & 0.281 & 0.071 & 1.842 \\
cellMCD & 0.559 & 0.222 & 0.120 & 1.084 \\
UGY95   & 0.453 & 0.245 & 0.178 & 1.183 \\
UHS88   & 0.499 & 0.204 & 0.169 & 0.934 \\
UHS88CC & 0.505 & 0.136 & 0.208 & 0.913 \\ 
\hline
\end{tabular}
\end{center}
\caption{Estimated parameters by several cellwise robust procedures when outliers size is $4$. True values are $\Phi = (0.5, 0.2, 0.2)$ and $\sigma = 1$.}
\label{tab:performance}
\end{table}

\begin{table}
\begin{center}
\begin{tabular}{lrrrrrr}
& \multicolumn{2}{c}{UGY95} & \multicolumn{2}{c}{UGY60} & \multicolumn{2}{c}{UHS88} \\
\hline
Obs. & NF  & F  & NF  & F  & NF  & F   \\
\hline
Clean       & 839 & 18 & 839 & 18 & 800 &  57 \\
Additive    &  45 & 98 &  44 & 99 &   0 & 143 \\
\hline
\end{tabular}
\end{center}
\caption{Perfomance of univariate filters. NF=Not Flagged and F=Flagged.}
\label{tab:flagged}
\end{table}

In conclusion, the issue of managing outliers in the analysis of time series is discussed. A formal definition of additive (AO) and innovative (IO) outliers is reviewed together with the relation between them and the identification problem. It is bliefly outlined the connection among the propagation of outliers and the cellwise contamination scheme which might be a key point to obtain robust procedures for time series. Performance of already available methods is  investigated in the particular case of additive outliers in autoregressive models with a simulated example.


\begin{thebibliography}{8}
\providecommand{\natexlab}[1]{#1}
\providecommand{\url}[1]{\texttt{#1}}
\expandafter\ifx\csname urlstyle\endcsname\relax
  \providecommand{\doi}[1]{doi: #1}\else
  \providecommand{\doi}{doi: \begingroup \urlstyle{rm}\Url}\fi

\bibitem[Agostinelli et~al.(2015)Agostinelli, Leung, Yohai, and
  Zamar]{agostinelli2015}
C.~Agostinelli, A.~Leung, V.J. Yohai, and R.H. Zamar.
\newblock Robust estimation of multivariate location and scatter in the
  presence of cellwise and casewise contamination.
\newblock \emph{TEST}, 24:\penalty0 441--461, 2015.

\bibitem[Danilov et~al.(2012)Danilov, Yohai, and Zamar]{danilov2012}
M.~Danilov, V.J. Yohai, and R.H. Zamar.
\newblock Robust esimation of multivariate location and scatter in the presence
  of missing data.
\newblock \emph{Journal of the American Statistical Association}, 107:\penalty0
  1178--1186, 2012.

\bibitem[Gervini and Yohai(2002)]{gervini2002}
D.~Gervini and V.J. Yohai.
\newblock A class of robust and fully efficient regression estimators.
\newblock \emph{The Annals of Statistics}, 30\penalty0 (2):\penalty0 583--616,
  2002.

\bibitem[Hannan and Rissanen(1982)]{hannan1982}
E.J. Hannan and J.~Rissanen.
\newblock Recursive estimation of mixed autoregressive-moving average order.
\newblock \emph{Biometrika}, 69\penalty0 (1):\penalty0 81--94, 1982.

\bibitem[Leung et~al.(2016)Leung, Zhang, and Zamar]{leung2016}
A.~Leung, H.~Zhang, and R.~Zamar.
\newblock Robust regression estimation and inference in the presence of
  cellwise and casewise contamination.
\newblock \emph{Computational Statistics \& Data Analysis}, 99:\penalty0 1--11,
  2016.

\bibitem[Raymaekers and Rousseeuw(2023)]{raymaekers2022}
J.~Raymaekers and P.J. Rousseeuw.
\newblock The cellwise minimum covariance determinant estimator.
\newblock \emph{Journal of the American Statistical Association}, 2023.
\newblock \doi{10.1080/01621459.2023.2267777}.

\bibitem[Raymaekers and Rousseeuw(2024)]{raymaekers2023}
J.~Raymaekers and P.J. Rousseeuw.
\newblock Challenges of cellwise outliers.
\newblock \emph{Econometrics and Statistics}, 2024.
\newblock \doi{https://doi.org/10.1016/j.ecosta.2024.02.002}.

\bibitem[Saraceno and Agostinelli(2021)]{saraceno2021}
G.~Saraceno and C.~Agostinelli.
\newblock Robust multivariate estimation based on statistical depth filters.
\newblock \emph{TEST}, pages 1--25, 2021.

\end{thebibliography}
\end{document}